\documentclass{aastex}
\usepackage{emulateapj5}
\usepackage{apjfonts}

\newcommand{\swas}{{\it SWAS}}
\newcommand{\water}{H$_2$O}
\newcommand{\wateriso}{H$_2^{18}$O}
\newcommand{\ceio}{C$^{18}$O}
\newcommand{\thco}{$^{13}$CO}

\newcommand{\hh}{H$_2$}

\journalinfo{{\rm Published in} The Astrophysical Journal, 539:L101--L105,
             {\rm 2000 August 20}}
\slugcomment{Received 2000 March 27; accepted 2000 June 21; published
             2000 August 16}

\shortauthors{WATER ABUNDANCE IN MOLECULAR CLOUD CORES}
\shorttitle{SNELL ET AL.}

\begin{document}

\title{Water Abundance in Molecular Cloud Cores}

\author{R. L. Snell\altaffilmark{1},
J. E. Howe\altaffilmark{1},
M. L. N. Ashby\altaffilmark{2},
E. A. Bergin\altaffilmark{2},
G. Chin\altaffilmark{3},
N. R. Erickson\altaffilmark{1},
P. F. Goldsmith\altaffilmark{4},
M. Harwit\altaffilmark{5},
S. C. Kleiner\altaffilmark{2},
D. G. Koch\altaffilmark{6},
D. A. Neufeld\altaffilmark{7},
B. M. Patten\altaffilmark{2},
R. Plume\altaffilmark{2},
R. Schieder\altaffilmark{8},
J. R. Stauffer\altaffilmark{2},
V. Tolls\altaffilmark{2},
Z. Wang\altaffilmark{2},
G. Winnewisser\altaffilmark{8},
Y. F. Zhang\altaffilmark{2},\\
and G. J. Melnick\altaffilmark{2}}

\altaffiltext{1}{Department of Astronomy, University of Massachusetts,
                 Amherst, MA 01003}
\altaffiltext{2}{Harvard-Smithsonian Center for Astrophysics, 60 Garden Street,
                 Cambridge, MA 02138}
\altaffiltext{3}{NASA Goddard Spaceflight Center, Greenbelt, MD 20771}
\altaffiltext{4}{National Astronomy and Ionosphere Center, Department of
                 Astronomy, Cornell University, Space Sciences Building,
                 Ithaca, NY 14853-6801}
\altaffiltext{5}{511 H Street SW, Washington, DC 20024-2725; also Cornell
                 University}
\altaffiltext{6}{NASA Ames Research Center, Moffett Field, CA 94035}
\altaffiltext{7}{Department of Physics and Astronomy, Johns Hopkins University,
                 3400 North Charles Street, Baltimore, MD 21218}
\altaffiltext{8}{I. Physikalisches Institut, Universit\"{a}t zu K\"{o}ln,
                 Z\"{u}lpicher Strasse 77, D-50937 K\"{o}ln, Germany}

\begin{abstract}

We present {\it Submillimeter Wave Astronomy Satellite} (\swas)
observations of the $1_{10}\rightarrow1_{01}$ transition of ortho-\water\
at 557 GHz toward 12 molecular cloud cores.  The water emission was
detected in NGC~7538, $\rho$ Oph A, NGC~2024, CRL~2591, W3, W3(OH), Mon
R2, and W33, and was not detected in TMC-1, L134N, and B335.  We also
present a small map of the \water\ emission in S140.  Observations of the
\wateriso\ line were obtained toward S140 and NGC~7538, but no emission
was detected.  The abundance of ortho-\water\ relative to \hh\ in the
giant molecular cloud cores was found to vary between $6\times10^{-10}$
and $1\times10^{-8}$.  Five of the cloud cores in our sample have previous
\water\ detections; however, in all cases the emission is thought to
arise from hot cores with small angular extents.  The \water\ abundance
estimated for the hot core gas is at least 100 times larger than in the
gas probed by \swas.  The most stringent upper limit on the ortho-\water\
abundance in dark clouds is provided in TMC-1, where the 3$\sigma$
upper limit on the ortho-\water\ fractional abundance is $7\times10^{-8}$.

\end{abstract}


\section{Introduction}

The {\it Submillimeter Wave Astronomy Satellite} (\swas) has detected spatially
extended \water\ emission associated with the quiescent dense molecular
gas in Orion and M17 \citep{sne00a,sne00b}.  Based on these observations
the relative abundance of ortho-\water\ was found to be surprisingly
small, ranging from only $1\times10^{-9}$ to $8\times10^{-8}$.  We present
here \swas\ observations of the ortho-\water\ transition at 557 GHz in
the three dark cloud cores TMC-1, L134N and B335 and in the nine giant
molecular cloud cores W3, W3(OH), Mon R2, $\rho$ Oph A, W33, CRL~2591,
S140, and NGC~7538.

Five of the cloud cores in our survey have had previously reported
water detections.  NGC~7538, S140, W3, and CRL~2591 were detected by
the {\it Infrared Space Observatory} in the ro-vibrational lines of
\water\ from the $\nu_2$ bending mode at 6 $\mu$m in absorption
against the dust continuum \citep{van96,hel96,van98}.  Water
emission has also been detected in NGC~7538, W3, and W3(OH) from the
$3_{13}\rightarrow2_{20}$ transition of \water\ at 183 GHz \citep{cer90}
and the $3_{13}\rightarrow2_{20}$ transition of \wateriso\ at 203 GHz
\citep*{gen96}. The 183 GHz and 203 GHz lines arise from transitions
with relatively high energies ($E_u/k \ga 200$ K) and thus are likely
produced in relatively warm gas.  The 6 $\mu$m absorption features can
arise from a range of gas temperatures, however the absorption seen
in these sources is thought to be dominated by relatively warm gas.
The relative abundance of \water\ in the warm gas has been estimated
to be $1\times10^{-6}$ to $1\times10^{-4}$.  \swas, on the other hand,
can observe the lowest energy rotational transition of ortho-\water\
permitting detection of \water\ emission from the more extended, cold
core gas.  We estimate the abundance of ortho-water in the relatively
cold ($T < 50$ K) core gas and compare the abundance results with those
for the warm gas and those for Orion and M17.

\section{Observations and Results}

The observations of \water\ and \wateriso\ reported here were obtained by
\swas\ during the period 1998 December to 2000 January.  The data were
acquired by nodding the satellite alternatively between the cloud core
and a reference position free of molecular emission.  Details concerning
data acquisition, calibration, and reduction with \swas\ are presented
in \citet{mel00}.  Observations of the $1_{10} \rightarrow 1_{01}$
transition of \water\ at a frequency of 556.936 GHz were obtained
toward the positions given in Table 1.  Total integration times for
these observations range from 17 hr for W3(OH) to 127 hr for L134N.

In addition, we obtained a small 10-point map of the S140 cloud core with
integration times per position of between 18 and 36 hr.  The center
positions of S140 and NGC~7538 were also observed in \wateriso\ at a
frequency of 547.676 GHz for 189 and 36 hr respectively.  The \swas\
beam is elliptical, and at the frequency of the water transitions has
angular dimensions of $3\farcm3 \times 4\farcm5$.  The data shown in
this paper are in antenna temperature units and are not corrected for
the measured \swas\ main beam efficiency of 0.90.  We also used the
Five College Radio Astronomy Observatory (FCRAO) 14~m telescope to map the
$J=1\rightarrow0$ transition of \ceio\ in an approximately $6\arcmin
\times 6\arcmin$ region in each core to provide an estimate of the gas
column density for use in our analysis of the \water\ data.

\begin{figure*}[t]
\epsscale{1.3}
\plotone{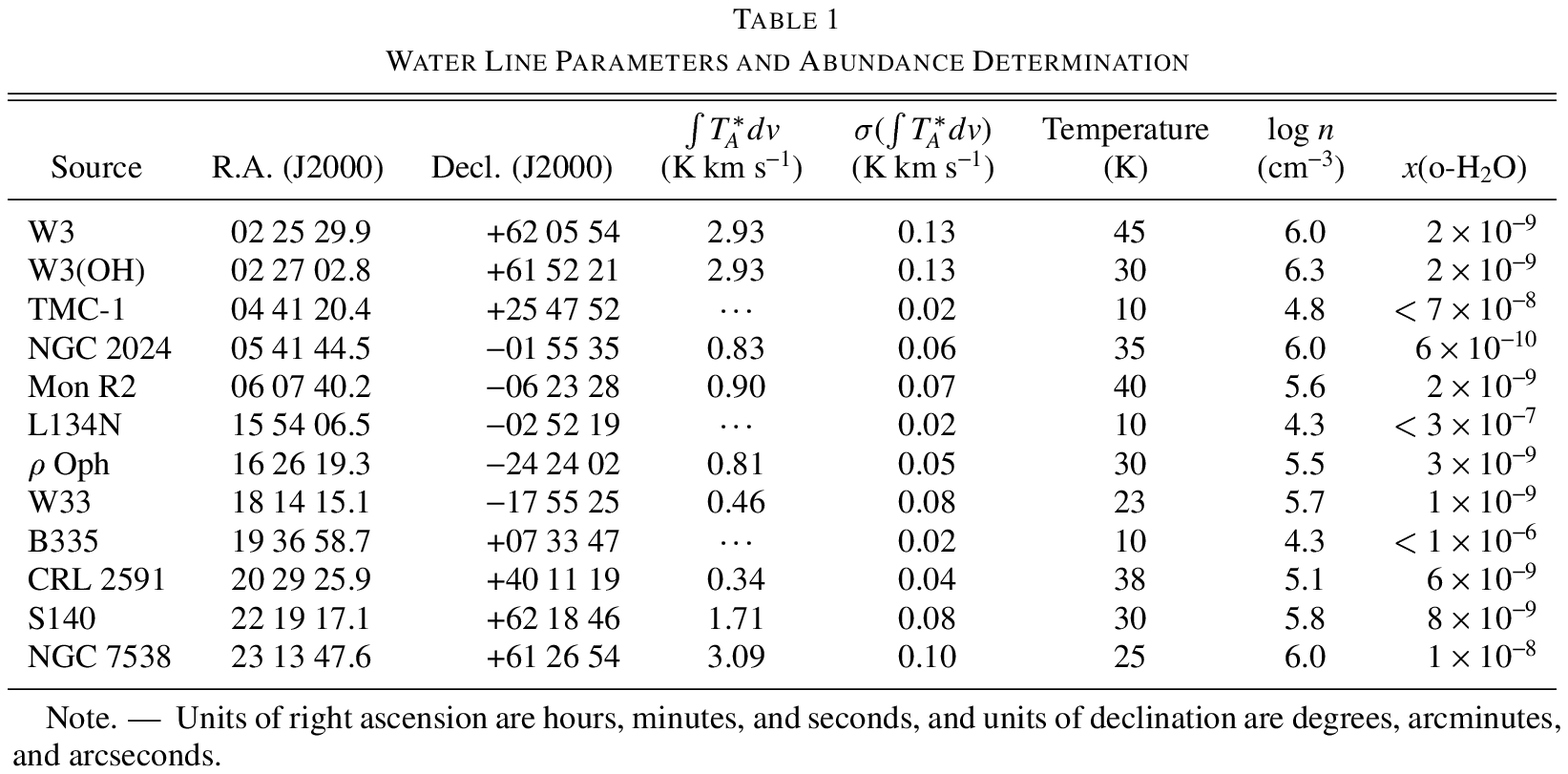}
\end{figure*}

\water\ emission was detected in all nine giant molecular cloud cores.
However despite deep integrations, we were unable to detect \water\
emission in any of the three dark cloud cores.  The \water\ spectra
obtained for six of the cloud cores are shown in Figure 1.  Spectra for
five of the remaining six cores are shown in \citet{ash00a}. The spectrum
of B335, where no \water\ emission was detected, is not shown. The \water\
line profiles for the giant molecular cloud cores show a wide variety
of shapes, ranging from relatively narrow single-peaked lines to broad,
strongly self-absorbed lines.  We have marked in Figure 1 the line
center velocity of the \thco\ $J = 5\rightarrow4$ emission (J. Howe et
al.\ 2000, in preparation) observed simultaneously with \swas.  In W3,
Mon R2, and W33 it is obvious that the \water\ line is self-absorbed,
since the peak of the \thco\ emission corresponds to a minimum in the
\water\ emission.  CRL~2591 is probably also self-absorbed, although in
this case the situation is somewhat more ambiguous.  More discussion about
the \water\ line profiles can be found in \citet{ash00a}.  The spectra
for $\rho$ Oph A and NGC~7538 presented in \citet{ash00a} show obvious
broad wings, suggesting that some of the \water\ emission probably arises
in associated molecular outflows.

A small map of the \water\ emission in S140 is shown in Figure
\hbox to\columnwidth{
2.  The emission toward the center of the core reveals a rela-
}

\vspace{1.5\baselineskip}
\centerline{\includegraphics[width=0.92\hsize,clip]{cores_h2o_fig1.ps}}
\figcaption{Spectra of the $1_{10}\rightarrow1_{01}$ transition of
ortho-\water\ from six cloud cores.  The line center velocity of the
\thco\ $J=5\rightarrow4$ transition observed simultaneously with \swas\
is indicated for W3, Mon R2, CRL~2591, and W33 by a dotted line. The
coordinates of each core are given in Table 1.
\label{fig1}}

\noindent
tively narrow ($\Delta V_{\rm FWHM}$ = 5.7 km s$^{-1}$), single-peaked
line.  \water\ emission was also detected at positions 3\farcm2 to the
north and south of center, however no emission was found east or west
of center.  The off-center spectra in general have rms noise of 0.02 K,
about 2 times larger than the center position.  During these observations
the long axis of the \swas\ beam was oriented approximately north-south,
thus explaining the stronger emission detected in these directions.

We failed to detect \wateriso\ emission in either S140 or NGC~7538.
The rms noise achieved in the NGC~7538 observation of 0.03 K was not
adequate to place a very useful limit on the \water/\wateriso\ ratio.
In S140, however, the \wateriso\ spectrum had an rms noise of only 0.01 K.
In S140 we used the line width and line center velocity derived from a
Gaussian fit to the \water\ line (see Fig. 2) to constrain a Gaussian
fit to the \wateriso\ spectra.  From this constrained fit we derived
a 1$\sigma$ upper limit to the integrated intensity of 0.022 K km
s$^{-1}$ leading to a 1$\sigma$ lower limit on the intensity ratio of
\water/\wateriso\ of 70.

\section{Determination of the Water Abundance}

We have analyzed the \water\ observations of the twelve cores

\vspace{1.5\baselineskip}
\centerline{\includegraphics[width=0.92\hsize,clip]{cores_h2o_fig2.ps}}
\figcaption{10-position map of the $1_{10}\rightarrow1_{01}$ transition
of ortho-\water\ obtained with \swas\ toward S140.  The spectra were
obtained on a regular grid spaced  by 3\farcm2 and are shown in their
correct relative positions on the sky.  Offsets in arcminutes relative
to the position of S140 given in Table 1 are indicated for each spectrum.
\label{fig2}}

\noindent
in the same manner as reported for Orion and M17 (Snell et al.\ 2000a,
2000b).  Before reviewing this procedure, it is worth noting that
the critical density for the 557 GHz transition of \water\ observed
by \swas\ is of order $10^{8}$ cm$^{-3}$, much higher than the average
density of gas in these cores.  Therefore the probability of collisional
deexcitation of this transition is very small, and photons produced
by collisional excitations will all escape the cloud core, although
they may be repeatedly absorbed and reemitted.  A linear relation thus
exists between the observed line integrated intensity and the product
of the density and the ortho-\water\ column density, independent of the
optical depth of this transition \citep{lin77,sne00a}.  The emission in
this case is effectively optically thin.  Although we did not use this
analytical expression to derive abundances, it provides insight into the
dependence of the water abundance on the physical properties of the cloud.

We estimate the relative ortho-\water\ abundance using a simple model of
the temperature, density, column density, and velocity dispersion for the
gas in each core.  The gas column density distribution was determined
from the \ceio\ maps, sampled at 44\arcsec\ intervals, assuming LTE
and a \ceio/\hh\ abundance ratio of $1.7\times10^{-7}$.  The \swas\
beam filling factor is effectively the convolution of the column density
distribution with the \swas\ beam pattern.  The velocity dispersion of
the gas along each line of sight was determined from the \ceio\ line width.
Another important parameter in our modeling is the gas density.  For five
sources (W3, W3(OH), $\rho$ Oph, W33, and Mon R2) we have strip maps
of the CS $J = 2\rightarrow1$ emission obtained at FCRAO and the $J =
5\rightarrow4$ and $J = 7\rightarrow6$ transitions obtained at KOSMA.
Most of this data is unpublished, although the results for the core
centers can be found in \citet{tro00}.  The CS emission was found to
be centrally peaked in these cores with detectable emission extending
at least several arcminutes from the center.  Modeling of these
data suggests that even though the column density decreases rapidly
away from the core center, the density remains relatively constant.
We used the published density results for L134N \citep{dic00}, TMC-1
\citep{pra97}, S140 \citep{sne84,zho94}, NGC~2024 \citep{sne84,sch91},
B335 \citep{zho93}, NGC~7538 \citep{plu97} and CRL~2591 \citep{car95,van99}
in our modeling.  For L134N, TMC-1, S140, and NGC~2024 these authors
have found that the density is relatively constant across the core.
Thus for many of the cores the high column density gas, which dominates
the \swas\ emission, can be roughly characterized by a single density.
We therefore assumed a constant density and temperature for the cores,
and the values used are summarized in Table 1.  Although a single density
cannot apply to all of the gas within the \swas\ beam, we believe that
the use of this approximation should not lead to serious errors in our
abundance determinations.

The collisional excitation of \water\ was computed using both the para- and
ortho-\hh\ collision rates with ortho-\water\ \citep*{phi96}, and assuming
a ratio of ortho- to para-\hh\ given by LTE at the temperatures given in
Table 1.  We did not include dust in our model, however the calculations
of \citet{ash00b} for S140 and $\rho$ Oph A indicate that the exclusion
of the dust continuum emission will have only a minor impact on the
determination of the water abundance.  We do note that in W33, the line
is clearly absorbing the dust continuum, and therefore some of the dust
continuum photons are converted into line photons.  For our analysis
we have used the total integrated intensity including the negative
contribution in the absorption feature, and thus, the integrated line
intensity should be representative of the total number of line photons.

We proceed in our analysis by guessing a relative ortho-\water\ abundance
for the core, then computing the emission distribution, and finally
convolving that distribution with the \swas\ beam.  The abundance
of \water\ is varied until the modeled integrated intensity agrees
with observations.  We assume a constant \water\ abundance across each
source. For positions with no detections, we used the 3$\sigma$ upper
limit on the integrated intensity to establish a 3$\sigma$ upper limit
on the \water\ abundance.  The abundances and/or limits found for each
core are given in Table 1.  For S140 and $\rho$ Oph A our results agree
well with the abundances derived by \citet{ash00b} using a more detailed
3-dimensional model of the density and temperature profile within these
sources.  The uncertainty in our derived abundances is dominated by the
uncertainties in the physical conditions, chiefly density.  The accuracy
of density determinations is of order a factor of three, thus leading to
abundance uncertainties of the same magnitude.  However, due to our use
of a single density to characterize these cores, the gas density at the
periphery of the cloud cores may be overestimated.  This does not have a
big impact on the abundance determination because the column density is
low in these regions and does not contribute substantially to the \water\
emission detected by \swas.  Nevertheless, the use of a single density
slightly biases our abundance determinations toward lower values.

In the giant molecular cloud cores the abundance of ortho-\water\
relative to \hh\ varies from $6\times10^{-10}$ to $1\times10^{-8}$.
The highest \water\ abundances are derived for NGC~7538 and S140.
The high \water\ abundance found for NGC~7538 is in part a consequence of
including emission arising from the molecular outflows, thus resulting
in an overestimate of the core abundance.  However, in S140, the line
profile is relatively narrow and it is unlikely that the core emission
is substantially contaminated by outflow emission.

We have also analyzed the mapping data for S140 using the same model
as used for the center position.  The detected \water\ emission at
positions 3\farcm2 north and 3\farcm2 south of center is consistent
with the predictions based on our cloud model using the abundance of
\water\ found for the core center.  The cloud core is extended to the
north-east, and the absence of \water\ emission toward the positions east
and north-east of the core center requires that either the abundance
of ortho-\water\ or the gas density be about 2 to 4 times smaller than
toward the core center.  The absence of extended emission in the higher
rotational transitions of CS \citep{sne84}, suggests that the lack of
\water\ emission is most likely explained by a lower density and not a
lower abundance.  The non-detections of \water\ emission in the remaining
positions is entirely consistent with our cloud model and the \water\
abundance reported in Table 1.

Although we believe that for most of the gas the \water\ emission is
effectively thin, it is possible that there are very small dense knots
of gas that are not.  In addition, a relatively thin layer of gas in an
extended envelope surrounding the core could have sufficient \water\
optical depth to scatter the line photons that arise from the core
\citep{sne00a}.  The \wateriso\ line should be less sensitive to these
problems since it is optically thin, and therefore a more reliable
estimator of the \water\ abundance.  We have used the \wateriso\
observations toward the S140 core to establish an upper limit to the
fractional abundance of ortho-\water.  Using the same physical model
described above, we derive a $3\sigma$ upper limit to the ortho-\wateriso\
abundance of $3\times10^{-10}$.  Assuming a ratio of H$_2$O/H$_2^{18}$O
of 500, this corresponds to a $3\sigma$ upper limit of $1\times10^{-7}$
for the abundance of ortho-\water\ relative to \hh.  Comparing this
result to our estimate of the \water\ abundance obtained from the more
common isotopic species, it is not surprising that we were unable to
detect \wateriso\ in this core.

We failed to detect \water\ emission from the three dark cloud cores
observed by \swas.  Although we achieved similar signal-to-noise ratio in
all three cores, the higher density and column density in TMC-1 permit
establishing a more stringent limit on the ortho-\water\ abundance. In
TMC-1, we find a $3\sigma$ upper limit on the fractional abundance of
ortho-\water\ of $7\times10^{-8}$.  However, even in TMC-1 the water
abundance limit is nearly two orders of magnitude larger than the typical
giant molecular cloud core.  The lower density and temperature in dark
cloud cores make detection of \water\ very difficult.

\section{Discussion and Summary}

\swas\ has detected thermal water emission from nine giant molecular cloud
cores which permit us to make the first estimate of the water abundance in
the cold, dense core gas.  The abundance of ortho-\water\ relative to \hh\
in these cloud cores varies from $6\times10^{-10}$ to $1\times10^{-8}$.
The relative ortho-\water\ abundance in the majority of these cores is
of order $1\times10^{-9}$, similar to the abundances found for M17SW
\citep{sne00b}.  In S140 the relative abundance of ortho-\water\ is
about ten times larger, closer to the average ortho-\water\ abundance for
Orion \citep{sne00a}.  In general the relative abundance of ortho-\water\
found in these cores is inconsistent with chemical equilibrium models
of well-shielded regions of clouds \citep*{ber98}, even considering the
most extreme uncertainties that might arise in our modeling procedure.
Possible explanations for this discrepancy are discussed in \citet{ber00}.

Water has been previously detected from the hot gas in W3 IRS~5,
W3(OH), CRL~2591, S140, and NGC~7538 IRS~1 via the 183 and 203 GHz
emission lines and the ro-vibrational absorption lines at 6 $\mu$m
\citep{van96,hel96,gen96,van98}.  It it worth noting that the \swas\
observation of W3 is centered about midway between IRS~4 and IRS~5,
less than 1\arcmin\ from IRS~5; and the \swas\ observation of NGC~7538
is centered about 1\arcmin\ south of IRS~1.  In both cases the regions
probed via the 6 $\mu$m, 183 GHz, and 203 GHz transitions are well
within the \swas\ beam.  The water emission and absorption in these
cores is thought to arise from hot cores surrounding young, massive
stars with angular extents of only a few arcseconds and gas temperatures
$\gtrsim$300 K \citep{van98,van99}.  Abundance estimates for water
in the hot cores range from $1\times10^{-6}$ to $1\times10^{-4}$,
several orders of magnitude greater than the abundance of water in the
surrounding core gas.  Even our limit on the \water\ abundance based
on \wateriso\ in S140 rules out the possibility that the large \water\
abundance found for the warm gas applies to the general cloud core.
At temperatures exceeding $\backsim$200-400 K, found in shocks and hot
cores, a series of neutral-neutral reactions rapidly converts nearly
all of the gas phase oxygen into water \citep{cha97,ber98}.  Thus, the
substantial difference predicted by the gas-phase chemistry models for
the water abundance in warm and cold gas is confirmed by our observations.

\swas, with a large beam size and the ability to observe a low excitation
line of water, is most sensitive to the water emission from the extended,
colder dense gas.  However, if the water abundance in the cold gas is
extremely small, then the hot cores may still contribute to the emission
seen by \swas.  If we assume that the brightness temperature of the
557 GHz emission from the hot core gas is 300 K (the same excitation
temperature as implied by the absorption lines seen by {\it Infrared
Space Observatory}) and that the core has an angular size of at most
2\arcsec, then the contribution to the observed \swas\ \water\ emission
is negligible in most sources, except CRL~2591.  Although \citet{tak99}
suggest that the CRL~2591 hot core has an angular diameter much less than
1\arcsec, the extremely weak emission detected in this source could have a
significant hot-core contribution.  Further information on the temperature
and angular size of the hot core is needed to evaluate this possibility.
If a significant fraction of the water emission detected by \swas\ in
CRL~2591 is attributed to the hot core, then the abundance of \water\
in the cold gas is even smaller than that quoted in Table 1.

Only upper limits to the ortho-\water\ abundance could be established
for the dark cloud cores.  If the abundance of water in these cores is
as small as that found for the giant molecular cloud cores, it will
be difficult to detect water emission in quiescent dark cloud cores.
The most stringent limit on the ortho-\water\ abundance is set in TMC-1,
and in this core the 3$\sigma$ upper limit on the \water\ abundance
is just marginally consistent with equilibrium gas-phase chemistry
predictions \citep{ber00}.  B335 is the site of a well-studied class
0 protostar, while neither L134N nor TMC-1 are actively forming stars.
Broad \water\ emission has been detected by \swas\ toward two other class
0 protostellar cores, L1157 and NGC~1333 IRAS~4 \citep{neu00}.  The \water\
emission in these sources clearly arises from the molecular outflows
associated with class 0 sources.  The relative abundance of ortho-\water\
in the outflow gas found by \citet{neu00} is of order $1\times10^{-6}$,
considerably larger than the limit set in the two non-star forming cores,
L134N and TMC-1.  \citet{neu00} suggests that the enhanced abundance is
due to shocks that convert a large fraction of the gas-phase oxygen into
water.  B335 also has an associated molecular outflow, with a outflow
mass between that of L1157 and NGC~1333 IRAS~4 \citep{mor89,neu00}.
The absence of \water\ emission in B335 is difficult to understand,
unless either the density of the outflow gas is much lower than in L1157
or NGC~1333 IRAS~4, or the shock velocities are sufficiently small that
\water\ formation by neutral-neutral reactions is inhibited.

\acknowledgements

This work was supported by NASA's \swas\ contract NAS5-30702 and NSF
grant AST97-25951 to the Five College Radio Astronomy Observatory.
R. Schieder and G. Winnewisser would like to acknowledge the generous
support provided by the DLR through grants 50 0090 090 and 50 0099 011.


\begin{thebibliography}{}

\bibitem[Ashby et al.(2000a)]{ash00a} Ashby, M. L. N., et al.\ 2000a, \apjl,
    539, L115 
\bibitem[Ashby et al.(2000b)]{ash00b} Ashby, M. L. N., et al.\ 2000b, \apjl,
    539, L119 
\bibitem[Bergin et al.(1998)Bergin, Melnick, \& Neufeld]{ber98} Bergin, E. A.,
    Melnick, G. J., \& Neufeld, D. A. 1998, \apj, 499, 777
\bibitem[Bergin et al.(2000)]{ber00} Bergin, E. A., et al. 2000, \apjl, 539,
    L129 
\bibitem[Carr et al.(1995)]{car95} Carr, J. S., Evans, N. J., II, Lacy, J. H.,
    \& Zhou, S. 1995, \apj, 450, 667
\bibitem[Cernicharo et al.(1990)]{cer90} Cernicharo, J., Thum, C., Hein, H.,
    John, D., Garcia, P., \& Mattioco, F. 1990, \aap, 231, L15
\bibitem[Charnley (1997)]{cha97} Charnley, S. B., 1997, \apj, 481, 396
\bibitem[Dickens et al.(2000)]{dic00} Dickens, J. E., Irvine, W. M., Snell,
    R. L., Bergin, E. A., Schloerb, F. P., Pratap, P., \& Miralles, M. P.
    2000, \apj, in press
\bibitem[Gensheimer et al.(1996)Gensheimer, Mauersberger, \& Wilson]{gen96}
    Gensheimer, P. D., Mauersberger, R., \& Wilson, T. L. 1996, \aap, 314, 281
\bibitem[Helmich et al.(1996)]{hel96} Helmich, F. P., et al.\ 1996, \aap, 315,
    L173
\bibitem[Linke et al.(1977)]{lin77} Linke, R. A., Goldsmith, P. F., Wannier,
    P. G., Wilson, R. W., \& Penzias, A. A. 1977, \apj, 214, 50
\bibitem[Melnick et al.(2000)]{mel00} Melnick G. J., et al.\ 2000, \apjl, 539,
    L77 
\bibitem[Moriarty-Schieven \& Snell(1989)]{mor89} Moriarty-Schieven, G. H., \&
    Snell, R. L. 1989, \apj, 338, 952
\bibitem[Neufeld et al.(2000)]{neu00} Neufeld, D. A., et al.\ 2000, \apjl, 539,
    L107 
\bibitem[Phillips et al.(1996)Phillips, Maluendes, \& Green]{phi96} Phillips,
    T. R., Maluendes, S., \& Green, S. 1996, \apjs, 107, 467
\bibitem[Plume et al.(1997)]{plu97} Plume, R., Jaffe, D. T., Evans, N. J., II,
    Mart\'{i}n-Pintado, J., \& G\'{o}mez-Gonz\'{a}lez, J. 1997, \apj, 476, 730 
\bibitem[Pratap et al.(1997)]{pra97} Pratap, P., Dickens, J. E., Snell, R. L.,
    Miralles, M. P., Bergin, E. A., Irvine, W. M., \& Schloerb, F. P. 1997,
    \apj, 486, 862
\bibitem[Schulz et al.(1991)]{sch91} Schulz, A., G\"{u}sten, R., Zylka, R.,
    Serabyn, E. 1991, \aap, 246, 570
\bibitem[Snell et al.(1984)]{sne84} Snell, R. L., Mundy, L. G., Goldsmith,
    P. F., Evans, N. J., II, \& Erickson, N. R. 1984, \apj, 276, 625
\bibitem[Snell et al.(2000a)]{sne00a} Snell, R. L., et al.\ 2000a, \apjl, 539,
    L93 
\bibitem[Snell et al.(2000b)]{sne00b} Snell, R. L., et al.\ 2000b, \apjl, 539,
    L97 
\bibitem[Trojan(2000)]{tro00} Trojan, C. 2000, Ph.D. thesis, Universit\"{a}t
    zu K\"{o}ln 
\bibitem[van der Tak et al.(1999)]{tak99} van der Tak, F. F. S., van Dishoeck,
    E. F., Evans, N. J., II, Bakker, E. J., \& Blake, G. A. 1999, \apj, 522,
    991
\bibitem[van Dishoeck(1998)]{van98} van Dishoeck, E. F. 1998, Faraday
    Discuss., 109, 31
\bibitem[van Dishoeck \& Helmich(1996)]{van96} van Dishoeck, E. F., \&
    Helmich, F. P. 1996, \aap, 315, L177
\bibitem[van Dishoeck et al.(1999)]{van99} van Dishoeck, E. F., et al.\ 1999,
    in The Universe as Seen by {\it ISO}, ed.\ P. Cox, \& M. F. Kessler,
    (ESA SP-427; Noordwijk: ESA/ESTEC), 437
\bibitem[Zhou et al.(1994)]{zho94} Zhou, S., Butner, H. M., Evans, N. J., II,
    G\"{u}sten, R., Kutner, M. L., \& Mundy, L. G. 1994, \apj, 428, 219
\bibitem[Zhou et al.(1993)]{zho93} Zhou, S., Evans, N. J., II, K\"{o}mpe, C.,
    \& Walmsley, C. M. 1993, \apj, 404, 232
\end{thebibliography}
\end{document}